\documentstyle[epsfig,12pt]{article}
\def\lsim{\mathrel{\mathpalette\fun <}}
\def\gsim{\mathrel{\mathpalette\fun >}}
\def\fun#1#2{\lower3.6pt\vbox{\baselineskip0pt\lineskip.9pt
\ialign{$\mathsurround=0pt#1\hfil##\hfil$\crcr#2\crcr\sim\crcr}}}

\newcommand{\be}{\begin{eqnarray}}
\newcommand{\ee}{\end{eqnarray}}

\begin{document}

\rightline{\Large{Preprint RM3-TH/99-6}}

\vspace{2cm}

\begin{center}

\Large{SEMI-INCLUSIVE ELECTRON SCATTERING OFF THE DEUTERON AND THE 
NEUTRON STRUCTURE FUNCTIONS\footnote{\bf To appear in the Proceedings of 
the International Workshop on {\em Exclusive and Semi-Exclusive Processes 
at High Momentum Transfer}, Jefferson Lab (Newport News, USA), May 20-22, 
1999.}}

\vspace{2cm}

\large{Silvano Simula}

\vspace{1cm}

\normalsize{Istituto Nazionale di Fisica Nucleare - Sezione Roma III\\
Via della Vasca Navale 84, I-00146 Roma, Italy} 

\vspace{1cm}

\abstract{The detection of slow nucleons in semi-inclusive electron scattering processes off the deuteron, $^2H(e, e'N)X$, is a very promising tool for the extraction of {\em model-independent} information on the neutron structure function $F_2^n(x, Q^2)$. The key point is the occurrence of a peculiar scaling property of the semi-inclusive cross section, which holds in the deep-inelastic regime as well as in the nucleon-resonance production regions. Moreover, the possibility to get unique information on the neutron asymmetry $A_1^n(x, Q^2)$ from polarised $^2\vec{H}(\vec{e}, e'N)X$ processes is illustrated.}

\end{center}

\newpage

\rightline{}

\newpage

\setcounter{page}{1}

\pagestyle{plain}

\section{INTRODUCTION}

Till now the investigation of the neutron structure function $F_2^n(x, Q^2)$ has been carried out by means of inclusive lepton scattering experiments off nuclear (typically deuteron) targets \cite{DATA,SLAC}. However, the unfolding of the neutron contribution from the inclusive cross section is known to lead to non-trivial ambiguities, which are mainly related to our non-precise knowledge of the nucleon-nucleon ($NN$) interaction and of the reaction mechanism (see Ref. \cite{SLAC} and references therein quoted). Moreover, the neutron asymmetry $A_1^n(x, Q^2)$ is still a poor-known quantity, particularly at large Bjorken $x$, where nucleon structure models might be tested (see Ref. \cite{isgur}).

A new way to get information on the neutron structure function has been recently proposed in Refs. \cite{SIM96,SIM98}. There it has been shown that the detection of slow nucleons in semi-inclusive electron scattering processes off the deuteron, $^2H(e, e'N)X$, is a very promising tool for the extraction of the neutron structure function in a {\em model-independent} way. In this contribution we recall the main results obtained in Refs. \cite{SIM96,SIM98} in case of unpolarised scattering and we briefly extend the analysis to the case of polarised $^2\vec{H}(\vec{e}, e'N)X$ processes.

\section{UNPOLARISED $^2H(e, e'N)X$ SCATTERING AND THE SPECTATOR SCALING}

Let us consider the so-called spectator mechanism in which, after virtual photon absorption by a nucleon $N_1$ in the deuteron, the recoiling nucleon $N_2$ is emitted and detected in coincidence with the scattered electron. The corresponding semi-inclusive cross section can be cast in the following form \cite{SIM96,SIM98} 
 \be
    {d^4 \sigma \over dE_{e'} ~ d\Omega_{e'} ~ dE_2 ~ d\Omega_2} = K M p_2 
    n^{(D)}(p_2) ~ {F_2^{N_1}(x^*, Q^2) \over x^*} ~ D^{N_1}(x, Q^2, 
    \vec{p}_2)
    \label{xsection}
 \ee
where $n^{(D)}(p_2)$ is the (non-relativistic) nucleon momentum distribution in the deuteron, $\vec{p}_2$ ($E_2$) is the momentum (energy) of the detected nucleon, $p_2 \equiv |\vec{p}_2|$, $Q^2 \equiv |\vec{q}|^2 - \nu^2$ is the squared four-momentum transfer, $x \equiv Q^2 / 2M\nu$ is the Bjorken scaling variable, $x^* \equiv Q^2 / (Q^2 + M_1^{*2} - M^2)$ and $M_1^*$ is the final invariant mass of the struck nucleon $N_1$, given through energy and momentum conservation by: $M_1^* = \sqrt{(\nu + M_D - E_2)^2 - (\vec{q} - \vec{p}_2)^2}$, with $M_D$ ($M$) being the deuteron (nucleon) mass. In Eq. (\ref{xsection}) $K \equiv (2M x^2 E_e E_{e'} / \pi Q^2)$ $(4 \pi \alpha^2 / Q^4)$ $\left [ 1 - y + (y^2 / 2) + (Q^2 / 4 E_e^2) \right ]$, with $y \equiv \nu / E_e$, and the quantity $D^{N_1}(x, Q^2, \vec{p}_2)$ depends both upon kinematical factors and the ratio $R_{L/T}^{N_1}$ of the longitudinal to transverse cross section off the nucleon (see Refs. \cite{SIM96,SIM98} for more details).

Let us introduce the following semi-inclusive response function \cite{SIM98}
 \be
     F^{(s.i.)}(x, Q^2, \vec{p}_2) & = & {1 \over \tilde{K}} {d^4
     \sigma \over dE_{e'} d\Omega_{e'} dE_2 d\Omega_2} \nonumber \\
     & = & M p_2 n^{(D)}(p_2) {F_2^{N_1}(x^*, Q^2) \over x^*} 
     \tilde{D}^{N_1}(x, Q^2, \vec{p}_2)
     \label{Fsi}
 \ee
where $\tilde{K}$ is a pure kinematical factor given by $\tilde{K} \equiv K ~ [D^{N_1}(x, Q^2, \vec{p}_2)]_{R_{L/T}^{N_1} = 0}$ and $\tilde{D}^{N_1}(x, Q^2, \vec{p}_2) =$ $D^{N_1}(x, Q^2, \vec{p}_2) ~ / ~ [D^{N_1}(x, Q^2, \vec{p}_2)]_{R_{L/T}^{N_1} = 0}$. In the Bjorken limit one should have  $R_{L/T}^{N_1} \to_{Bj} 0$ apart from logarithmic $QCD$ corrections; thus, $\tilde{D}^{N_1} \to_{Bj} 1$ and 
 \be
    F^{(s.i.)}(x, Q^2, \vec{p}_2) \to_{Bj} M p_2 n^{(D)}(p_2) ~ 
    {F_2^{N_1}(x^*)\over x^*}
    \label{Bjlimit}
 \ee
where $F_2^{N_1}(x^*) \to_{Bj} F_2^{N_1}(x^*, Q^2)$. Therefore, in the Bjorken limit and at fixed values of $p_2(\equiv |\vec{p}_2|)$ the semi-inclusive response function (\ref{Fsi}) does not depend separately upon $x$ and the nucleon detection angle $\theta_2$ ($cos\theta_2 \equiv \vec{p}_2 \cdot \vec{q} / p_2 q$), but it depends only upon the variable $x^* \to_{Bj} x / (2 - z_2)$, with $z_2 = [E_2 - p_2 cos\theta_2] / M$ being the light-cone momentum fraction of the detected nucleon. In what follows, we will refer to the variable $x^*$ and the quantity $F^{(sp)}(x^*, Q^2, p_2)$, given explicitly by 
 \be
      x^* & = & {Q^2 \over Q^2 + (\nu + M_D - E_2)^2 - (\vec{q} - 
      \vec{p}_2)^2 - M^2}
     \label{xsp} \\
     F^{(sp)}(x^*, Q^2, p_2) & = & M p_2 n^{(D)}(p_2) ~ F_2^{N_1}(x^*, Q^2) 
     / x^*
    \label{Fsp}
 \ee
as the {\em spectator scaling} variable and function, respectively.

A natural question is whether and to what extent the spectator scaling can hold as well at finite values of $Q^2$. To this end, Eq. (\ref{Fsi}) has been calculated for various kinematical conditions corresponding to $Q^2 = 1$ and $10 ~ (GeV/c)^2$; the Bjorken variable $x$ and the nucleon detection angle $\theta_2$ have been varied in the range $0.20 \div 0.95$ and $10^o \div 170^o$, respectively (for sake of simplicity, the polar angle $\phi_2$ has been chosen equal to $0$). As for the nucleon structure function, the parameterization of the $SLAC$ data of Ref. \cite{SLAC}, containing $R_{L/T}^{N_1} \simeq 0.18$, has been adopted. The results of the calculations of Eq. (\ref{Fsi}), performed at $p_2 = 0.3 ~ GeV/c$, are reported in Fig. 1 and compared with the corresponding values of the spectator-scaling function (\ref{Fsp}). It can clearly be seen that the spectator scaling is fulfilled within few $\%$ in the {\em deep inelastic scattering} ($DIS$) regime as well as in {\em nucleon-resonance production} regions. This result is due to the fact that the spectator variable $x^*$ collects final electron and nucleon kinematics corresponding to the same value of $M_1^*$; therefore, the spectator scaling does not depend upon the occurrence of the Bjorken scaling of the nucleon structure function.

\begin{figure}[htb]

\centerline{\epsfxsize=16cm \epsfig{file=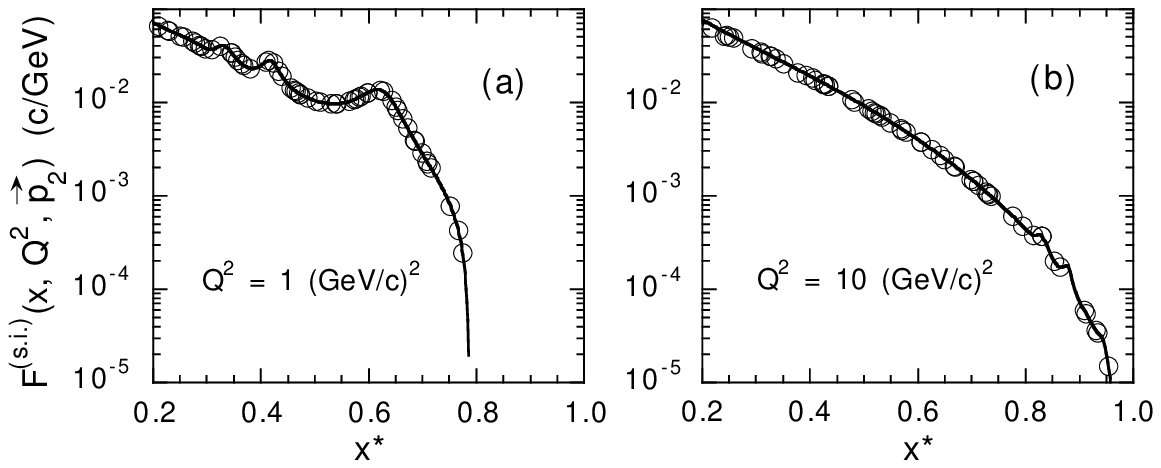}}

{\small {\noindent Figure 1. The response function $F^{(s.i.)}(x, Q^2, \vec{p}_2)$ (Eq. (\ref{Fsi})) for the process $^2H(e, e'p)X$ versus the spectator-scaling variable $x^*$ (Eq. (\ref{xsp})) calculated at $Q^2 = 1 ~ (GeV/c)^2$ (a) and  $Q^2 = 10 ~ (GeV/c)^2$ (b) for $p_2 = 0.3 ~ GeV/c$. The solid line is the spectator-scaling function (\ref{Fsp}).}}

\vspace{0.5cm}

\end{figure}

In the spectator-scaling function (\ref{Fsp}) all the nuclear effects are contained in the nucleon momentum distribution $n^{(D)}(p_2)$. At $p_2 \lsim 0.3 ~ GeV/c$ the dependence of $n^{(D)}(p_2)$ upon the nuclear force model is quite negligible, because at large internucleon distances the $NN$ interaction is almost totally governed by the one-pion-exchange contribution. Therefore, in the spectator-scaling regime it is possible to extract the nucleon structure function with small uncertainty due to nuclear effects. In particular, the process $^2H(e, e'p)X$ can provide information on the neutron structure function, while the reaction $^2H(e, e'n)X$, involving the proton structure function, can be used for a consistency check.

Within the spectator scaling regime the measurement of the semi-inclusive cross section both for the $^2H(e, e'p)X$ and $^2H(e, e'n)X$ processes allows to investigate two spectator-scaling functions, involving the same nuclear part, $M p_2 n^{(D)}(p_2)$, and the neutron and proton structure functions, respectively. Since moreover $R_{L/T}^n \simeq R_{L/T}^p$ (as suggested by recent $SLAC$ data analyses \cite{RLT_SLAC}), at fixed values of $p_2$ both the nuclear part and the factor $\tilde{D}^{N_1}(x, Q^2, \vec{p}_2)$ cancel out in the ratio of semi-inclusive cross sections (\ref{xsection}), yielding 
 \be
    R^{(s.i.)}(x, Q^2, \vec{p}_2) \equiv {d^4 \sigma[^2H(e,e'p)X]
    \over d^4 \sigma[^2H(e,e'n)X]} \to {F_2^n(x^*, Q^2) \over
    F_2^p(x^*, Q^2)}
    \label{ratio}
 \ee
and, therefore, the neutron to proton structure function ratio can directly be inferred from the ratio of semi-inclusive cross sections. Note that with respect to the response function $F^{(s.i.)}(x, Q^2, \vec{p}_2)$ the ratio $R^{(s.i.)}(x, Q^2, \vec{p}_2)$ exhibits a more general scaling property, for at fixed $Q^2$ it does not depend separately upon $x$ and $\vec{p}_2$, but only on $x^*$. This implies that any $p_2$-dependence of the semi-inclusive ratio (\ref{ratio}) would be a signature for off-shell deformations of the nucleon structure function (see Ref. \cite{SIM96}).  Finally, note that in the quasi-elastic scattering regime the spectator scaling analysis can be applied to extract {\em model-independent} information on the elastic-peak contribution to the nucleon structure functions (see Ref. \cite{SIM98}).

Since the spectator scaling is a peculiar feature of the spectator mechanism, it can be in principle violated by different reaction mechanisms. In Refs. \cite{SIM96,SIM98} the nucleon emission arising from the target fragmentation of the struck nucleon, which is thought to be responsible for the production of slow hadrons in $DIS$ processes, has been investigated at high $Q^2$ adopting the factorization approach described in Ref. \cite{SIM95}. The results of the calculations of the semi-inclusive ratio (\ref{ratio}), performed at $Q^2 = 10 ~ (GeV/c)^2$, are reported in Fig. 2 and compared with the free neutron to proton structure function ratio given in Ref. \cite{SLAC}. It can clearly be seen that: ~ i) only at $x^* \lsim 0.4$ fragmentation processes can produce relevant violations of the spectator scaling, as it is expected from the fact that the diquark remnant has a light-cone momentum fraction which decreases as $x$ increases; ~ ii) {\em backward kinematics} appear to be the most appropriate conditions for the extraction of the neutron to proton structure function ratio. 

\begin{figure}[htb]

\vspace{0.5cm}

\centerline{\epsfxsize=16cm \epsfig{file=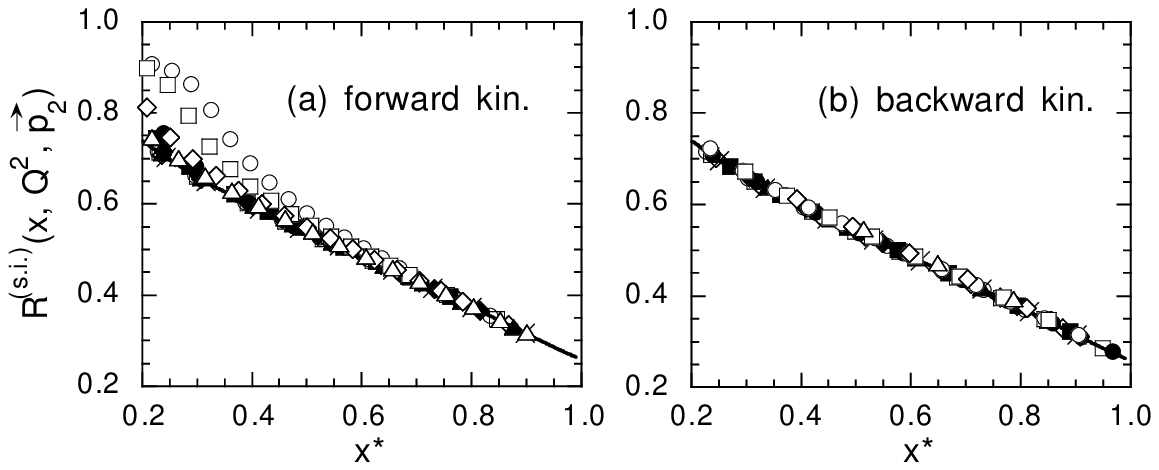}}

{\small {\noindent Figure 2. The semi-inclusive ratio $R^{(s.i.)}(x, Q^2, \vec{p}_2)$ (Eq. (\ref{ratio})) versus the spectator-scaling variable $x^*$ (Eq. (\ref{xsp})), calculated at $Q^2 = 10 ~ (GeV/c)^2$ and $p_2 = 0.1, 0.3, 0.5 ~ GeV/c$ by adding to the spectator mechanism the contribution arising from the target fragmentation of the struck nucleon. The various markers correspond to different values of the nucleon detection angle $\theta_2$; forward ($\theta_2 < 90^o$) and backward ($\theta_2 > 90^o$) nucleon emissions are shown in (a) and (b), respectively. The solid line is the free neutron to proton structure function ratio of Ref. \cite{SLAC}.}}

\vspace{0.5cm}

\end{figure}

Violations of the spectator scaling can be expected also from the distortions of the energy and angular distributions of the detected nucleon due to the reinteraction of the debris produced by the fragmentation of the struck nucleon with the recoiling spectator nucleon. Estimates of these final state interactions in the $DIS$ regime have been discussed in Refs. \cite{SIM96,SIM98}, where it has been argued that backward nucleon emission is not significantly affected by forward-produced hadrons. However, at low $Q^2$ ($\sim$ few $(GeV/c)^2$) pion production may lead to reinteraction effects with the recoiling spectator nucleon, if both particles are emitted backward in the lab frame. For sake of simplicity let us consider the process of virtual photon absorption on a nucleon initially at rest in the lab frame, producing a $\Delta(1232)$ resonance which subsequently decays into a nucleon plus a pion. The momentum of the nucleon and the pion in the center-of-mass ($CM$) of the resonance is given from energy-momentum conservation by $p_{CM} = \{ (M_{\Delta}^2 - M^2 + m_{\pi}^2)^2 / 4M_{\Delta}^2 - m_{\pi}^2\}^{1/2} \simeq 0.23 ~ GeV/c$. The boost velocity from the $CM$ to the lab frame is $V_q = q / \sqrt{M_{\Delta}^2 + q^2}$, with $q = |\vec{q}|$ being the three-momentum transfer. The component of the pion momentum in the lab frame along the direction of $\vec{q}$ can be easily evaluated and shown to be proportional to $(V_q - V_{\pi})$, where $V_{\pi} = p_{CM} / \sqrt{m_{\pi}^2 + p_{CM}^2} \simeq 0.85$. Therefore, the condition for having pions emitted only forward in the lab frame is $V_q \geq V_{\pi}$, which implies $q \gsim 2 ~ GeV/c$. Since in the lab frame $M + \nu = \sqrt{M_{\Delta}^2 + q^2}$, one gets $Q^2 \gsim 2 ~ (GeV/c)^2$. Thus, it can be safely argued that for $Q^2 \gsim 1 ~ (GeV/c)^2$ the final state interactions of the recoiling nucleon emitted in the backward hemisphere should drop quickly with increasing $Q^2$ and become almost totally negligible for $Q^2 \gsim$ few $(GeV/c)^2$. 

\section{POLARISED $^2\vec{H}(\vec{e}, e'N)X$ SCATTERING}

Let us now extend our analysis to the case of $^2\vec{H}(\vec{e}, e'N)X$ processes, in which both the target and the incident electron beam are polarised. In what follows we will limit ourselves to the Bjorken limit, since this suffices to illustrate the relevant features of the slow nucleon production. We start by defining the semi-inclusive deuteron asymmetry $A_1^{D(s.i.)}$ in terms of the helicity components of the semi-inclusive cross section, viz. $A_1^{D(s.i.)} = (\sigma_{3/2}^{(s.i.)} - \sigma_{1/2}^{(s.i.)}) / (\sigma_{3/2}^{(s.i.)} + \sigma_{1/2}^{(s.i.)})$, which in the Bjorken limit can be cast in the following form
 \be
    A_1^{D(s.i.)}(x, \vec{p}_2) = 2x {g_1^{D(s.i.)}(x, \vec{p}_2) \over 
    F_2^{D(s.i.)}(x, \vec{p}_2)}
    \label{A1D}
 \ee
where $g_1^{D(s.i.)}(x, \vec{p}_2)$ and $F_2^{D(s.i.)}(x, \vec{p}_2)$ are polarised and unpolarised semi-inclusive deuteron structure functions, respectively. Within the spectator mechanism one easily gets
 \be
    g_1^{D(s.i.)}(x, \vec{p}_2) & = & {1 \over z_1} g_1^{N_1}({x \over z_1})
    \Delta n^{(^2H)}(\vec{p}_2) [1 + {p_2 \over M} cos(\theta_2)]
    \nonumber \\
    F_2^{D(s.i.)}(x, \vec{p}_2) & = & z_1 F_2^{N_1}({x \over z_1})
    n^{(^2H)}(p_2)
    \label{g1DF2D}
 \ee
where $z_1 = 2 - z_2 \simeq 1 + p_2 cos(\theta_2) / M + O(p_2^2 / M^2)$ is the light-cone momentum fraction of the struck nucleon $N_1$. In Eq. (\ref{g1DF2D}) the deuteron structure is encoded in the functions $n^{(^2H)}(p_2)$ and $\Delta n^{(^2H)}(\vec{p}_2)$. In terms of the non-relativistic deuteron S- and D-wave functions $u(p)$ and $w(p)$, one has (cf. Ref. \cite{weise})
 \be
     n^{(^2H)}(p) & = & u^2(p) + w^2(p)
     \nonumber \\
     \Delta n^{(^2H)}(\vec{p}) & = & u^2(p) + {3cos^2(\theta) - 1 \over 
     \sqrt{2}} u(p)w(p) + [{3 \over 2} cos^2(\theta) - 1] w^2(p)
     \label{wf}
 \ee
Using Eqs. (\ref{A1D}-\ref{wf}), the semi-inclusive deuteron asymmetry $A_1^{D(s.i.)}(x, \vec{p}_2)$ acquires in the spectator mechanism a simple factorized form, viz.
  \be
    A_1^{D(s.i.)}(x, \vec{p}_2) = A_1^{N_1}(x^*) D(\vec{p}_2)
    \label{A1sp}
  \ee
where $A_1^{N_1}(x^*) = 2x^* g_1^{N_1}(x^*) / F_2^{N_1}(x^*)$ is the asymmetry of the struck nucleon, $x^* = x / (2 - z_2)$ is the spectator variable and $D(\vec{p}_2)$ is the depolarisation factor arising from the D-wave of the deuteron, namely
 \be
    D(\vec{p}_2) = {1 \over 1 + {w^2(p_2) \over u^2(p_2)}}  \left\{ 1 + 
   {3cos^2(\theta_2) - 1 \over \sqrt{2}} {w(p_2)  \over u(p_2)} + [{3 \over 
   2} cos^2(\theta_2) - 1] {w^2(p_2) \over u^2(p_2)}  \right\} ~~~~
    \label{depol}
 \ee
Thus, the semi-inclusive deuteron asymmetry (\ref{A1sp}) does not exhibit the spectator scaling property in the variable $x^*$, because of the explicit dependence of the depolarisation factor (\ref{depol}) on the nucleon detection angle $\theta_2$. A remarkable sensitivity of $D(\vec{p}_2)$ to the specific value of $\theta_2$ and to the $NN$ interaction model (at least for $p_2 \gsim 0.2 ~ GeV/c$) is illustrated in Fig. 3. 

\begin{figure}[htb]

\vspace{0.5cm}

\parbox{9cm}{\epsfxsize=8.75cm \epsfig{file=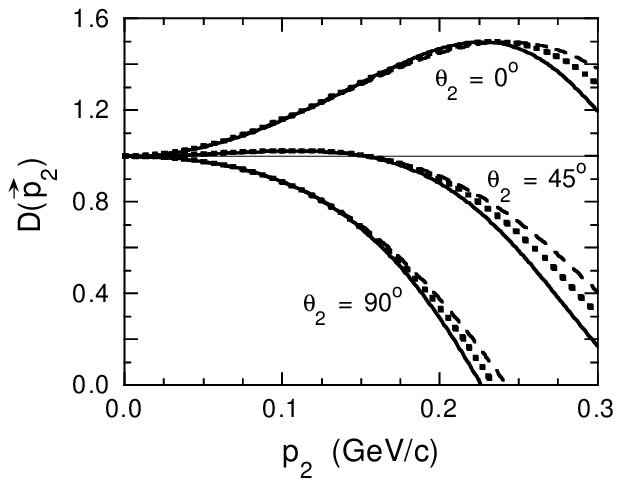}} \ $~~$ \ \parbox{6.25cm}{\small \noindent Figure 3. The depolarisation factor (\ref{depol}) versus the nucleon momentum $p_2$ for various values of the nucleon detection angle $\theta_2$. The solid, dashed and dotted lines correspond to the use of the $RSC$ \cite{pot}$^{(a)}$, Paris \cite{pot}$^{(b)}$ and Bonn \cite{pot}$^{(c)}$ models of the $NN$ interaction, respectively.}

\vspace{0.5cm}

\end{figure}

However, from Eq. (\ref{A1sp}) the spectator scaling is fulfilled by the ratio of semi-inclusive deuteron asymmetries, viz.
 \be
      R_A^{D(s.i.)}(x, \vec{p}_2) \equiv {A_1^{D(s.i.)}[^2\vec{H}(\vec{e}, 
     e'p)X] \over A_1^{D(s.i.)}[^2\vec{H}(\vec{e}, e'n)X] } \to {A_1^n(x^*) 
     \over A_1^p(x^*)}
     \label{ratioA1}
 \ee
so that the ratio $R_A^{D(s.i.)}(x, \vec{p}_2)$ does not depend separately on $x$ and $\vec{p}_2$, but only on $x^*$. The value of the scaling function is now directly the neutron to proton asymmetry ratio $A_1^n(x^*) / A_1^p(x^*)$, which can be therefore extracted from deuteron data in a {\em model-independent} way.

\indent Before closing this Section we point out that in the process $^2\vec{H}(\vec{e}, e'n)X$ the deuteron asymmetry$A_1^{D(s.i.)}(x, \vec{p}_2)$ involves the proton asymmetry $A_1^p(x^*)$ and the depolarization factor $D(\vec{p}_2)$. Since the former can be obtained using hydrogen targets, the depolarisation factor (\ref{depol}) can be extracted at low momenta from the reaction $^2\vec{H}(\vec{e}, e'n)X$. In this way unique information on the D- to S-wave ratio in the deuteron can be obtained and compared with theoretical predictions based on different $NN$ interaction models, provided $p_2 \simeq 0.2 \div 0.3 ~ GeV/c$ (see Fig. 3).

\section{CONCLUSIONS}

The production of slow nucleons in semi-inclusive (polarised and unpolarised) electron scattering off the deuteron may exhibit a peculiar scaling property, the spectator-scaling, which can be used to extract {\em model-independent} information on the neutron structure functions. Namely, the neutron structure function $F_2^n$, the neutron to proton structure function ratio $F_2^n / F_2^p$ and the neutron to proton asymmetry ratio $A_1^n / A_1^p$ can be obtained directly from the semi-inclusive cross section data in the kinematical regions corresponding to deep inelastic scattering ($HERA$-like kinematics) as well as to nucleon-resonance production ($JLAB$-like kinematics), provided slow nucleons (i.e., with momentum of the order of $0.1 \div 0.2 ~ GeV/c$) are detected in the backward hemisphere.

\section*{Acknowledgments} The author gratefully acknowledges Gunther Piller for supplying him with the parameterizations of the deuteron wave function used in this work.


\section*{References}

\begin{thebibliography}{99}

\bibitem{DATA} J.J. Aubert et al., Nucl. Phys. B {\bf 293}, 740 (1987). A.C. Benvenuti et al., Phys. Lett. B {\bf 237}, 599 (1990). P. Amaudruz et al., Nucl. Phys. B {\bf 371}, 3 (1992).

\bibitem{SLAC} A. Bodek and J.L. Ritchie, Phys. Rev. D {\bf 23}, 1070 (1981). L.W. Whitlow et al., Phys. Lett. B {\bf 282}, 745 (1992).

\bibitem{isgur} N. Isgur, Phys. Rev. D {\bf 59}, 341033 (1999).

\bibitem{SIM96} S. Simula, Phys. Lett. B {\bf 387}, 245 (1996); e-print archive nucl-th 9608053, Proc. of the Workshop on {\em Future Physics at HERA} (DESY, September '95 - May '96), eds. G. Ingelman, A. de Roeck and R.K. Klanner, DESY (Hamburg, 1996), p. 1058.

\bibitem{SIM98} S. Simula, Nucl. Phys. A {\bf 631}, 602c (1998).

\bibitem{RLT_SLAC} S. Dasu et al., Phys. Rev. D {\bf 49}, 5641 (1994).

\bibitem{SIM95} C. Ciofi degli Atti and S. Simula, Few Body Systems {\bf 18}, 55 (1995). S. Simula, Few Body Systems Suppl. {\bf 9}, 466 (1995).

\bibitem{weise} S.A. Kulagin et al., Phys. Rev. C {\bf 52}, 932 (1995). For a recent review see: G. Piller and W. Weise, preprint TUM/T39-99-11, e-print archive hep-ph9908230.

\bibitem{pot} (a) V. Reid, Ann. Phys.  (N.Y.) {\bf 50}, 411 (1968). (b) M. Lacombe et al., Phys. Rev. C {\bf 21}, 861 (1980). (c) R. Machleidt et al., Phys. Rep. {\bf 149}, 1 (1987).

\end{thebibliography}
\end{document}